\documentclass[aps,pra,superscriptaddress,longbibliography,twocolumn, floatfix]{revtex4-1}

\usepackage[english]{babel}
\usepackage{amsmath}
\usepackage{amssymb}
\usepackage{graphicx}
\usepackage{color}
\usepackage{xspace}

\newcommand{\lco}{$\mathrm{La_2CuO_4}$\xspace}
\newcommand{\lcod}{$\mathrm{La_2CuO_{4+\delta}}$\xspace}
\newcommand{\lsco}{$\mathrm{La_{2-x}Sr_xCuO_4}$\xspace}
\newcommand{\lscod}{$\mathrm{La_{2-x}Sr_xCuO_{4+\delta}}$\xspace}
\newcommand{\dlco}[1]{$\delta$-LCO$_{#1}$\xspace}
\newcommand{\musr}{$\mu$SR\xspace}
\newcommand{\lem}{LE-$\mu$SR\xspace}
\newcommand{\ie}{\emph{i.e.}\xspace}

\begin{document}
 
\title{Superconductivity drives magnetism in $\boldsymbol\delta$-doped $\mathbf{La_2CuO_4}$}

\date{\today}

\author{A.~Suter}
\email[corresponding author:~]{andreas.suter@psi.ch}
\affiliation{Laboratory for Muon Spin Spectroscopy, Paul Scherrer Institute, CH-5232 Villigen PSI, Switzerland}
\author{G.~Logvenov}
\author{A.V.~Boris}
\author{F.~Baiutti}
\author{F.~Wrobel}
\affiliation{Max Planck Institute for Solid State Research, Heisenbergstrasse 1, 70569 Stuttgart, Germany}
\author{L.~Howald}
\affiliation{SLS, Paul Scherrer Institute, CH-5232 Villigen PSI, Switzerland}
\author{E.~Stilp}
\affiliation{Laboratory for Muon Spin Spectroscopy, Paul Scherrer Institute, CH-5232 Villigen PSI, Switzerland}
\affiliation{Materials for Energy Conversion, Empa, CH-8600 D\"{u}bendorf, Switzerland}
\author{Z.~Salman}
\author{T.~Prokscha}
\affiliation{Laboratory for Muon Spin Spectroscopy, Paul Scherrer Institute, CH-5232 Villigen PSI, Switzerland}
\author{B.~Keimer}
\affiliation{Max Planck Institute for Solid State Research, Heisenbergstrasse 1, 70569 Stuttgart, Germany}

\begin{abstract}
The understanding of the interplay between different orders in a solid is a key challenge in highly correlated electronic
systems. In real systems this is even more difficult since disorder can have a strong influence on the subtle balance
between these orders and thus can obscure the interpretation of the observed physical properties. Here we present a study 
on $\delta$-doped \lco (\dlco{N}) superlattices. By means of molecular beam epitaxy whole $\mathrm{LaO_2}$-layers were periodically
replaced through $\mathrm{SrO_2}$-layers providing a charge reservoir, yet reducing the level of disorder typically present
in doped cuprates to an absolut minimum. The induced superconductivity and its interplay with the antiferromagnetic order
is studied by means of low-energy \musr. We find a quasi-2D superconducting state which couples to the antiferromagnetic order
in a non-trivial way. Below the superconducting transition temperature, the magnetic volume fraction increases strongly.
The reason could be a charge redistribution of the free carriers due to the opening of the superconducting gap which is possible
due to the close proximity and low disorder between the different ordered regions.
\end{abstract}


\maketitle

The copper oxide based high-temperature superconductors (cuprates) exhibit rich and
complex physics \cite{keimer_quantum_2015}. Strong electron correlations drive the parent 
compounds into an insulating, antiferromagnetic ground state. Upon sufficiently high doping 
of the copper oxide planes by electrons or holes, superconductivity appears. Still, even 
for doping levels were the highest superconducting transition temperature, $T_c$, 
is reached, short range antiferromagnetic correlations sustain. In some cuprates, the competition 
between superconducting and magnetic orders causes a tendency towards electronic phase separation, 
especially on the underdoped side of the phase diagram. The phase coexistence of superconductivity 
and antiferromagnetic stripe order in the $\mathrm{La_{2-x-y}}M_{\rm y}\mathrm{Sr_xCuO_4}$
family was observed at finite temperatures by neutron scattering for $M=\mathrm{Nd}$ 
\cite{tranquada_evidence_1995} and \musr/NMR for $M=\mathrm{Eu}$ \cite{klauss_spin_2004}.
Subsequent intense theoretical efforts showed
(\cite{corboz_competing_2014} and references therein) that within the $t-J$ model, there
is close competition between uniform $d$-wave superconductivity and various stripe states and
the real ground state is very susceptible to disorder. One source of disorder in the cuprates
are the dopant atoms, which is adding another level of complexity \cite{andersen_disorder-induced_2010}.
In this respect, superoxygenated \lcod \cite{wells_incommensurate_1997,mohottala_phase_2006} is 
an interesting family. There the excess oxygen is intercalating in a self-organized manner into 
the structure of antiferromagnetic and superconducting regions \cite{fratini_nature_2010} quite 
remarkably so that the magnetism and superconductivity set in at the same temperature, independent 
of Sr content and characteristic of optimally doped oxygen-stoichiometric \lscod \cite{udby_measurement_2013}.
Furthermore, the concomitant magnetic propagation vector remains consistent with that of the stripe 
ordered cuprates. 

\begin{figure*}[h]
 \centering
 \includegraphics[width=0.8\textwidth]{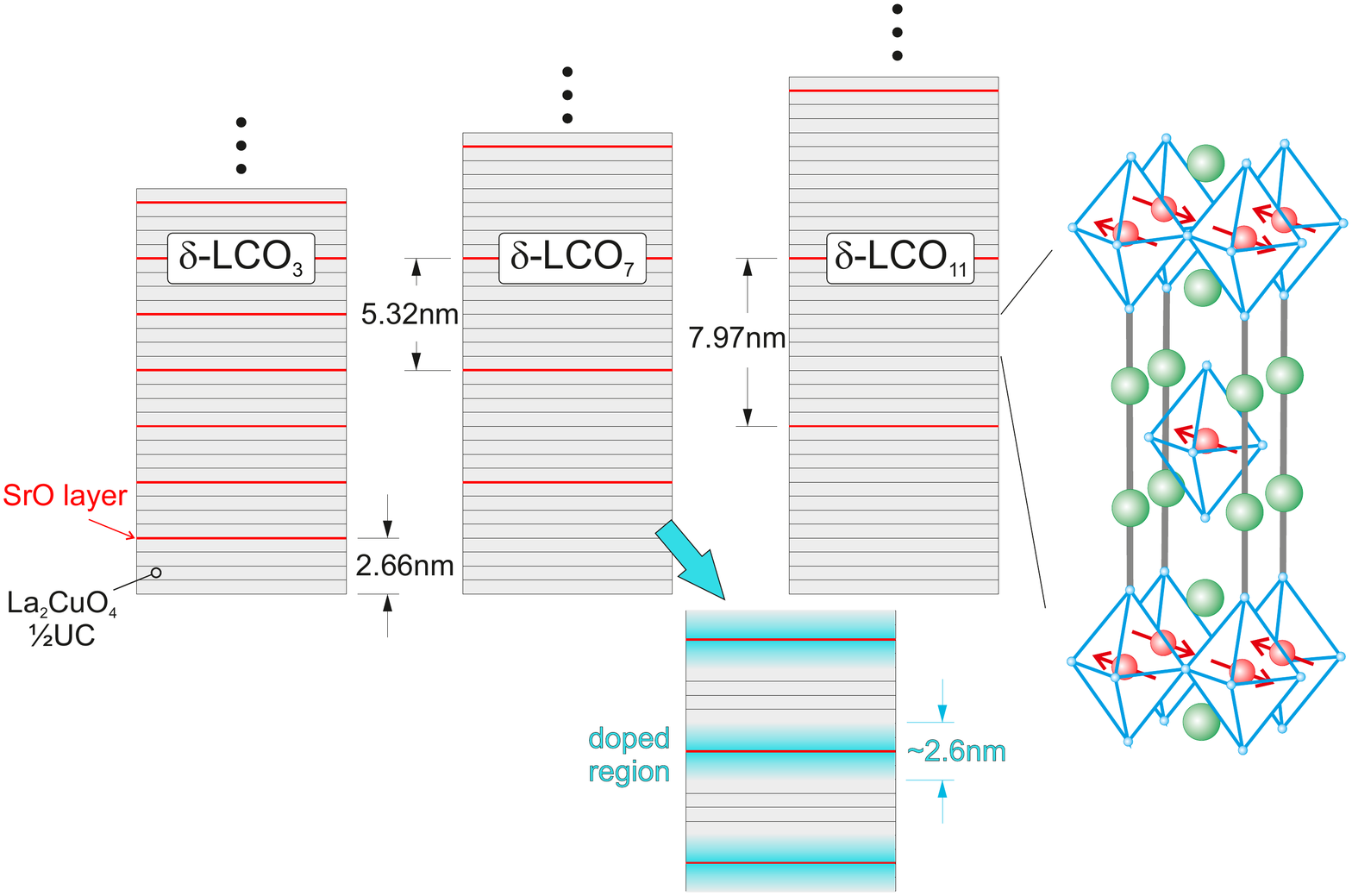}
 \caption{Sketch of delta doped \lco, \dlco{N}. Starting from \lco, a
          superlattice is formed by replacing single layers of LaO with SrO planes.
          The general formula can be written as: $R \times [$ SrO-LaO-CuO$_2 + N
          \times$ LaO-LaO-CuO$_2]$, \ie the natural counting is given in half 
          crystallographic unit cells. $R$ is adjusted such that the overall 
          thickness of the \dlco{N} superlattices is about 40~nm.
          The negatively charged interface region around the SrO-layer will lead
          to a layered charge distribution throughout the superlattice,
          as depicted with the light blue layers. An in-depth study about the
          structure and charge distribution within the \dlco{N} 
          superlattices is found in Ref.\cite{baiutti_high-temperature_2015}.}\label{fig:dLCO-Sketch}
\end{figure*}

In this paper we demonstrate a novel approach to dope \lco. Rather than randomly substituting lanthanum by strontium, 
which leads to micro-scale disorder, we replace single planes of LaO with SrO dopant planes
using atomic layer-by-layer molecular beam epitaxy \cite{baiutti_high-temperature_2015,wang_ACS_ApplMaterInterfaces_2016}. 
This allows a much better control over the disorder compared to bulk \lscod and, 
at the same time, gives another degree of freedom, namely the separation of the charge reservoirs. In this way 
a system on the mesoscopic scale can be engineered, allowing to tune the interplay between
superconducting and antiferromagnetic ground states. Figure \ref{fig:dLCO-Sketch} depicts a 
sketch for a selection of such superlattices which we call $\delta$-doped \lco. The distance
between SrO dopant layers can be labeled $N$ which is the number of half-unit-cells separating
them, and hence we will abbreviate this family by \dlco{N}. 

Utilizing low-energy muon spin rotation techniques, we find a non-trivial enhancement of the magnetic volume 
fraction below the superconducting transition of the \dlco{N} superlattices in striking resemblance to
bulk superoxygenated \lscod. Furthermore, it is shown that the superfluid density of \dlco{N} is in-line with the Uemura 
relation \cite{uemura_systematic_1988}, namely that the superfluid density is anomalously small 
and proportional to $T_c$ on the underdoped side.

\begin{figure*}[h]
 \centering
 \includegraphics[width=\textwidth]{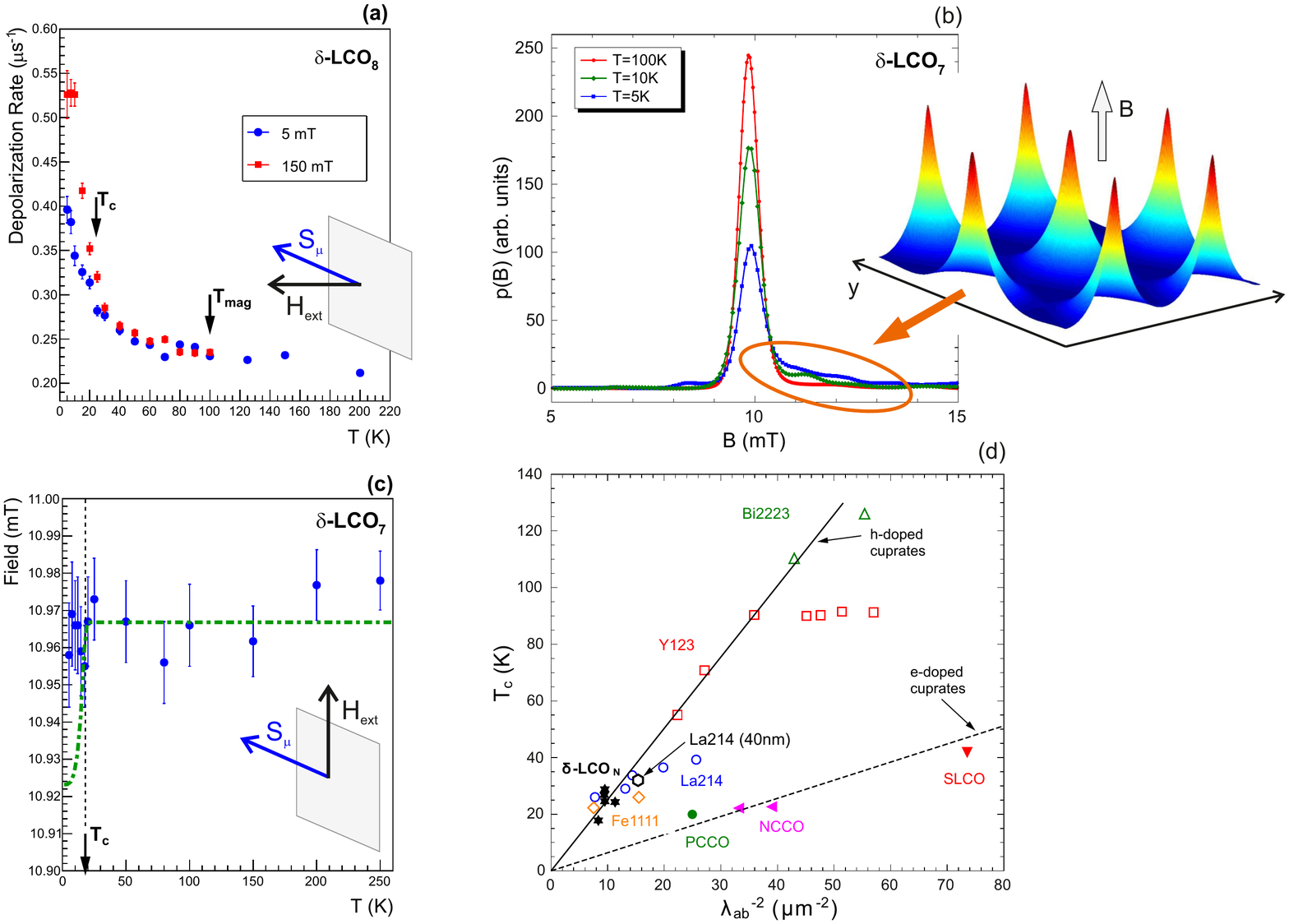}
 \caption{In (a), the muon depolarization rate, $\sigma(T)$, as function of the temperature 
          is shown. It is obtained from field cooling measurements with the applied magnetic 
          field, $H_{\rm ext}$, perpendicular to the \dlco{N} superlattice film axis. The 
          magnetic field probability distribution ($z$-components $\|$ to $H_{\rm ext}$) of 
          the vortex state is presented in (b). (c) shows the measured magnetic field in the Meissner 
          state. The green dash-dotted line shows the expected field dependence in the 
          Meissner state taking into account the $\lambda_{\rm L}$ obtained from the vortex 
          state. (d) shows the Uemura plot with the \dlco{N} results and an optimally doped 
          \lsco film with a thickness of 40~nm. Data of the other systems are from 
          Ref.\cite{luetkens_field_2008} and references there in.}\label{fig:dLCO-SC}
\end{figure*}

\lem allows to study internal magnetic field distributions of any material
\cite{morenzoni_nano-scale_2004}, and thus is very well suited to investigate
systems with a complex interplay between magnetic and superconducting ground
states. By tuning the implantation energy of the positive muon, the stopping
range can be varied between 5 and 300~nm (see also supplementary S.2). For this
study an implantation energy $E_{\rm impl}$ was chosen such that the full
muon beam stops in the center of the superlattice. The full stopping distribution 
can be found in the supplementary material. In order to obtain information about 
the superconducting state it is possible either to study the vortex state or the 
Meissner state. From measurements in the vortex state the magnetic field 
distribution is provided. For a regular vortex lattice, the second moment of the 
magnetic field distribution is proportional to the muon depolarization rate, $\sigma(T)$, 
and directly related to the magnetic penetration depth $\lambda(T)$ as (see 
Ref.\cite{brandt_flux_1988})

\begin{equation}\label{eq:sigma_sc}
  \left(\frac{\sigma_{\rm sc}(T)}{\gamma_\mu}\right)^2 = 0.00371\,
\frac{\Phi_0^2}{\lambda(T)^4},
\end{equation}

\noindent where $\gamma_\mu$ is the muon gyromagnetic ratio, $\sigma_{\rm sc} =
\sqrt{\sigma(T)^2 - \sigma(T>T_c)^2}$, $\Phi_0 = 2.067\cdot 10^{-15} (\mathrm{T m}^2)$ 
is the flux quantum. Figure \ref{fig:dLCO-SC} (a) shows the temperature dependence of 
$\sigma$ in the vortex state, and (b) presents the magnetic field probability 
distribution ($z$-components) of the vortex state given by the Fourier transform
of the muon spin polarization function (Supplementary S.2.1.2). The marked high field 
shoulder is typical for a regular vortex lattice. Since the film thickness, 
$d \simeq 40$~nm, is small compared to the London penetration depth, $\lambda_{\rm L}$,
$\lambda(T)$ in Eq.(\ref{eq:sigma_sc}) represents an effective magnetic
penetration depth \cite{clem_two-dimensional_1991}. The relation between them is
approximately given by $\lambda_{\rm L}^2(T) = c_0 \lambda(T)\cdot d$, with
$c_0 = 1/2$. Measurements on optimally doped $\mathrm{La_{2-x}Sr_xCuO_4}$ with
$d=40$~nm were scaled such that we obtained the bulk data, resulting in a $c_0 =
4.3$. We chose this factor to estimate $\lambda_{\rm L}$ for the superlattices. The
Uemura plot in (d) shows, that the \dlco{N} superlattices are in line with the
hole doped cuprates. Measurements in the Meissner state (zero field cooled, $H_{\rm ext} < H_{c1}$)
should show a corresponding magnetic field shift as depicted by the dash-dotted line in 
Fig.\ref{fig:dLCO-SC} (c). The absence of the Meissner state demonstrates
that superconductivity is layered in nature and likely localized around the charged
SrO-layers.

In metal-insulator superlattices of the form $R \times [3 \times
\mathrm{La_{1.55}Sr_{0.45}CuO_4} + N \times \mathrm{La_2CuO_4}]$ 
the charge transfer effects throughout the superlattices was modeled
quantitatively \cite{loktev_model_2008,suter_superconductivity_2012}. This is possible since
the chemical potential as function of Sr doping in \lsco has been experimentally 
determined \cite{ino_chemical_1997}. The result shows that superconducting layers 
along the interfaces form with an extend of about 1 UC. As for the \dlco{N}
superlattices, the Josephson coupling in the vortex state breaks down (field
geometry as in Fig.\ref{fig:dLCO-SC} (a)), and the Meissner state is suppressed (as in
Fig.\ref{fig:dLCO-SC} (c)). These findings are further supported by the temperature 
dependence of $\sigma(T)$ which does not follow the expected behavior $\sigma(T) 
\propto [1 - (T/T_c)^r]$, $r \simeq 2\ldots 6$. 

The situation is very reminiscent to the case of intercalated Bi2212 and Bi2202
\cite{baker_tuning_2009} where the interlayer spacing between adjacent
CuO$_2$-layers was tuned by intercalating guest molecules. Above a critical separation 
the Josephson coupling between adjacent layers is getting too weak and only the 
dipole-dipole interaction remains to align the pancake vortices. The $\sigma$ 
versus $T$ behavior found there is essentially identical to what is shown in 
Fig.\ref{fig:dLCO-SC} (a). The superconducting state of the \dlco{N} superlattices 
can be summarized such that superconducting layers are forming rather localized at around 
the SrO-layers. The distance between these quasi-2D superconducting layers ranges 
from $\sim 2.6$~nm for \dlco{3} up to $\sim 7.9$~nm for \dlco{11}, thus the Josephson 
coupling between layers is essentially suppressed and only dipolar interaction between 
vortices can stabilize the vortex lattice. Therefore the superconducting ground state is 
extremely anisotropic. A very recent infrared spectroscopy study of charge dynamics in \dlco{N} 
confirms that the superconducting state in this system is essentially two-dimensional \cite{boris_infrared_dlco}. 

\begin{figure*}[h]
 \centering
 \includegraphics[width=0.8\textwidth]{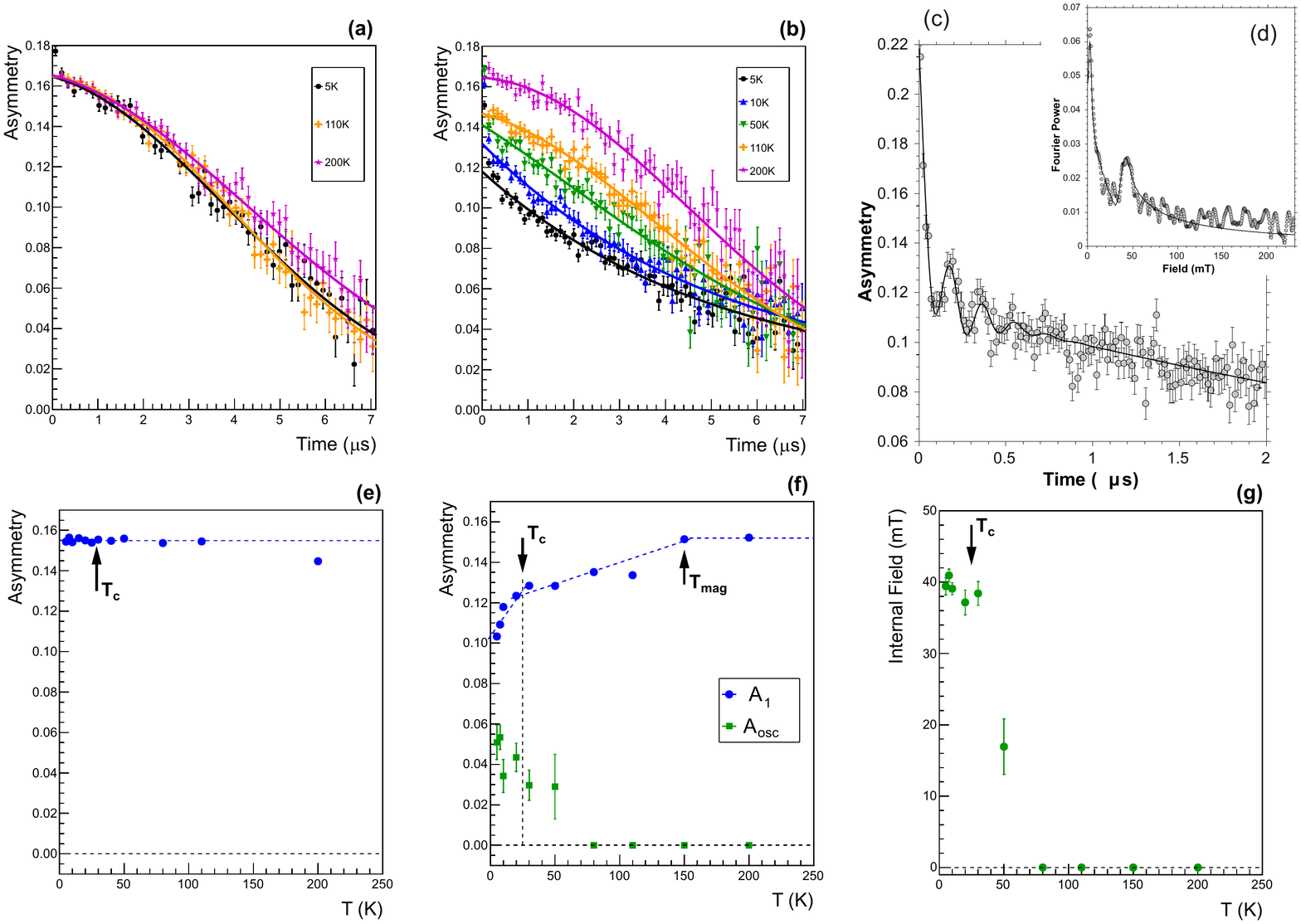}
 \caption{Zero field \lem data for $\delta$-LCO$_3$ and $\delta$-LCO$_{11}$. The measured 
          asymmetry, $A(t)$, is proportional to the muon spin polarization $P(t)$. (a) and (b) 
          show asymmetry time spectra of the \dlco{3} and \dlco{11} superlattice, respectively,
          measured at various temperatures. (e) and (f) show the initial asymmetry, $A(t=0)$,
          as function of temperature for the \dlco{3} and \dlco{11} superlattice, respectively.
          In (c) the short time asymmetry spectrum measured at $T=5$~K of \dlco{11} is presented, 
          where a clear spontaneous zero field precession is visible, with its Fourier transform 
          depicted in (d).           
          In (g) the temperature dependence of the internal magnetic field at the muon stopping site 
          is given. The corresponding precession amplitudes are found in (f).}\label{fig:dLCO-ZF}
\end{figure*}

\musr is a well-established method to study magnetic systems \cite{reotier_muon_1997}. Reasons 
are that the ground state can be studied in zero applied magnetic field, and a sensitivity of about 
$10^{-3}\mu_{\rm B}$ per unit cell is reached. The top panels of Fig.\ref{fig:dLCO-ZF} (a-c) show 
the time evolution of the muon spin asymmetry, $A(t) = A_0\, P(t)/P(0)$. $A_0$ is the instrumental 
asymmetry and $P(t)$ the muon decay asymmetry (see also supplementary 
material S.2). For the \dlco{3} superlattice $A(t)$ shows a Gaussian like time evolution typical 
for a paramagnetic state where the loss of the polarization is solely governed by the dephasing 
of the muon spin ensemble due to the quasi static nuclear magnetic dipole fields \cite{yaouanc_muon_2010}. The 
very weak temperature dependence of $A(t)$ is an indication of the gradual slowing down of high 
frequency short range magnetic correlation still present in the system. In Fig.\ref{fig:dLCO-ZF}~(e)
the temperature dependence of the initial asymmetry, $A(t=0)$, is presented which stays constant in 
the whole observed temperature range. These zero field results show that a SrO-layer separation of 
$\sim 2.6$~nm is close enough to fully suppress the AFM ground state of the \lco layers due to charge 
transfer. Essentially, \dlco{3} is behaving as a metal with short range AFM correlations.

\dlco{11} shows a drastically different behavior. The full time spectra shown in Fig.\ref{fig:dLCO-ZF}~(b)
change from an initially Gaussian like behavior at high temperature, towards an exponential one
at low temperature. At short times and low enough temperature, spontaneous zero field precession is
found (see Fig.\ref{fig:dLCO-ZF} (c)). This shows that \dlco{11}, differently to \dlco{3}, undergoes 
an antiferromagnetic transition. To be able to quantify the changes in the asymmetry spectra, the
following zero field fit model was assumed:

\begin{eqnarray}\label{eq:zf}
  A(t) &=& \left. A_0 P(t)/P(0) \right|_{\rm ZF} = \nonumber \\
       & & A_1 \left(\frac{1}{3} + \frac{2}{3} \left\{ 1 - (\Delta\cdot t)^2 \right\} \exp\left[ -\frac{1}{2} (\Delta\cdot t)^2 \right] \right) \,e^{-\lambda t} + \nonumber \\
       & & A_{\rm osc}\, j_0(\gamma_\mu B_{\rm int} t + \phi)\, e^{-(\sigma_{\rm osc} t)^2/2} + A_{\rm bkg}.
\end{eqnarray}

\noindent Since the muon stopping distribution is covering the whole superlattice (see the supplementary 
material), the asymmetry spectrum, $A(t)$, will be a superposition of muons experiencing a paramagnetic
surrounding (close to the SrO doping layers) and muons stopping in an antiferromagnetic surrounding (far 
from the SrO layers). The first term describes the paramagnetic response of the sample. $\Delta$ is the 
width of the magnetic field distribution due to nuclear dipoles and $\lambda$ is describing the slowing down
of high frequency short range magnetic correlations. The second term describes the regions which are 
antiferromagnetically ordered. The zero field precession signal is well described by a zero-order 
spherical Bessel function. The last term, $A_{\rm bkg}$, describes a background signal due to muons 
not stopping in the sample. For a more detailed discussion of Eq.(\ref{eq:zf}) see the supplementary 
material. The value of the internal magnetic field  $B_{\rm int}$ is a very sensitive measure of the doping level 
in \lsco \cite{borsa_staggered_1995,savici_muon_2002}. We find $B_{\rm int}(T\to 0) = 40(2)$~mT which allows to estimate
an upper doping level in the antiferromagnetic regions of $x<0.01$. Furthermore, this value shows that
the full electronic Cu moment of about $0.64\,\mu_{\rm B}$ is present in the antiferromagnetic state.
The zero field time spectra and temperature dependencies of the asymmetries of the \dlco{N}, $N=7,8,9$ are 
found in the supplementary materials. The loss of the temperature dependent paramagnetic asymmetry $1-A_1(T)/A_0$ 
reflects the growth of the magnetic volume fraction. Its behavior is rather surprising as can be seen in 
Fig.\ref{fig:dLCO-ZF}~(f). At about $T=150$~K, $A_1(T)$ starts to gradually decrease towards lower temperature. 
At $T_c$ a clear trend change can be observed, with a substantially faster increase of the magnetic volume fraction.
 
\begin{figure*}[h]
 \centering
 \includegraphics[width=0.8\textwidth]{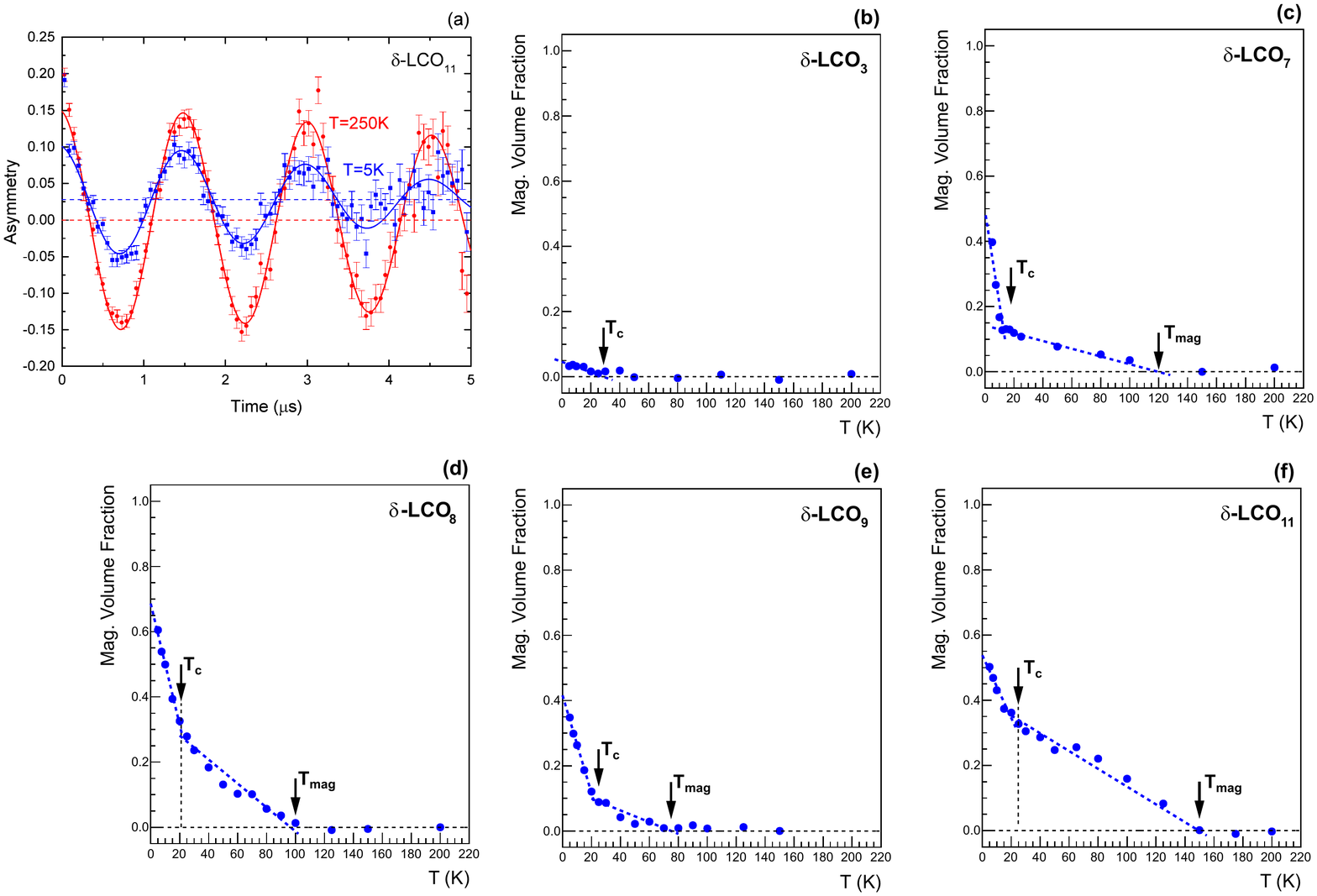}
 \caption{(a) Weak transverse field asymmetry time spectra for \dlco{11}, measured in a field of $\mu_0 H_{\rm ext}=5$~mT. 
          The red data set is measured in the paramagnetic phase at $T=250$~K, whereas the blue data set is measured
          at $T=5$~K. The low temperature data show an asymmetry offset, $A_{\rm L}$, which demonstrates that
          a fraction of the muons are stopping in a magnetic surrounding. (b)-(f) show the magnetic volume fractions, 
          $f_{\rm M}(T)$. All measured \dlco{N} superlattices show a clear change in slope at $T_c$, \ie the
          magnetic regions increase faster when the adjacent metallic layers become superconducting.}\label{fig:dLCO-MagVolFrac}
\end{figure*}

In order to quantify this effect, weak transverse field measurements (wTF) were carried out which allow to 
measure the magnetic volume fraction, $f_{\rm M}$, in an efficient and precise manner. The long-lived oscillation
amplitude in the wTF asymmetry represents muons in a non- or paramagnetic environment. 
Fig.\ref{fig:dLCO-MagVolFrac}~(a) shows typical wTF measurements in an applied field of $\mu_0 H_{\rm ext} = 5$~mT. The
data were fitted to

\begin{eqnarray}
 A(t) &=& \left. A_0 P(t)/P(0)\right|_{\rm wTF} = \nonumber \\
   & & A_{\rm T} e^{-(\sigma t)^2/2} \cos(\gamma_\mu [\mu_0 H_{\rm ext}] t + \phi) +
   A_{\rm L} \cos(\phi).
\end{eqnarray}

\noindent The magnetic volume fraction is given by $f_{\rm M} = 1 - A_{\rm T}/A_0$.
For all para- and diamagnetic states $A_{\rm L} \equiv 0$. Therefore, the finite value of $A_{\rm L}$ found below
$T_{\rm M}$, the $T=5$~K value is depicted in Fig.\ref{fig:dLCO-MagVolFrac} (a), clearly demonstrates the presence 
of a magnetic ground state. The low-temperature magnetic volume fraction allows to estimate
the superconducting layer thickness. Assuming that the superlattices are laterally homogeneous, with no stripe-like 
electron patterns within the superconducting layer, a magnetic and 
superconducting layer thickness can be estimated, as presented in Tab.\ref{tab:mag-sc-thickness}. It shows that the upper
limit for the superconducting layer thickness $d_{S}(0) \simeq\, $2-4~nm, as sketched in Fig.\ref{fig:dLCO-Sketch}. 
This value is consistent with the dopant profile in \dlco{N} as measured by high-resolution and analytical scanning 
transmission electron microscopy \cite{wang_ACS_ApplMaterInterfaces_2016}, and hence it is not too surprising that
\dlco{3} shows only marginal signs of magnetism, since $d_{N} \lesssim d_{S}(0)$.

\begin{table}[h]
 \centering
 \begin{tabular}{c|c|c|c|c|c|c|c|c} 
  N & $f_{\rm M}(T_c)$ & $f_{\rm M}(0)$ & $d_{N}$    & $d_{\rm M}(T_c)$ & $d_{\rm M}(0)$ & $d_{S}(0)$ & $T_c$  & $\lambda_{\rm ab}^{-2}$ \\
    &                  &                & (nm)       & (nm)             &  (nm)          & (nm)       & (K)    & ($\mu\mathrm{m}^{-2}$) \\ \hline\hline
  3 & 0                & 0              & 2.64       & 0                & 0              & 2.64       & 28.8   & 9.5(3) \\
  7 & 0.13             & 0.5            & 5.28       & 0.69             & 2.64           & 2.64       & 17.8   & 8.4(3) \\
  8 & 0.27             & 0.7            & 5.94       & 1.6              & 4.16           & 1.78       & 24.3   & 11.3(4) \\
  9 & 0.10             & 0.4            & 6.60       & 0.66             & 2.64           & 3.96       & 26.8   & 9.4(2) \\
 11 & 0.32             & 0.54           & 7.92       & 2.53             & 4.28           & 3.64       & 24.6   & 9.5(3) 
 \end{tabular}
 \caption{Estimates of the magnetic and superconducting thicknesses. The first and second column gives the magnetic 
          volume fraction at $T_c$ and zero temperature respectively. The superlattice repetition length is 
          $d_{N}=(N+1)\cdot\mathrm{UC}/2$, with $\mathrm{UC}=1.32$~nm. The magnetic layer thickness is therefore defined as 
          $d_{\rm M}(T) = f_{\rm M}(T)\cdot d_{N}$. An \emph{upper limit} for the superconducting layer thickness
          is thus $d_{S}(0) = d_{N} - d_{\rm M}(0)$. The last column gives the $T_c$'s of the superlattices.}\label{tab:mag-sc-thickness}
\end{table}

A closer look a the temperature dependence of $f_{\rm M}(T)$ reveals a rather unusual behavior. 
Typically, $f_{\rm M}(T)$ shows a sharp upturn at $T_{\rm M}$ as found in various copper- and iron-based
superconductors \cite{savici_muon_2002,udby_measurement_2013,bernhard_muon_2012,guguchia_negative_2014}. 
In contrast, for all \dlco{N}, $f_{\rm M}(T)$ increases very gradually, almost linearly, when lowering 
the temperature. However, at exactly $T_c$ there is a clear trend change, $df_{\rm M}/dT$ is strongly increasing. 
As shown in the supplementary material S.4, this behavior is also present when applying the external magnetic field, 
$H_{\rm ext}$, parallel to the superlattice layers, thus ruling out that the observed effect is related to the
formation of a vortex lattice in the superconducting state. This observation suggests that the magnetic and superconducting
ground states are coupled. 

In Ref.\cite{kivelson_thermodynamics_2001} the authors discuss, in the context of stripe formation, 
the coupling between incommensurate antiferromagnetic and superconducting order in terms of the 
thermodynamics of fluid mixtures. They confirm that $f_{\rm M}(T)$ may grow in the superconducting state, 
albeit not giving a microscopic explanation of the simultaneous onset of magnetism and superconductivity, 
$T_{\rm M} \approx T_{\rm c}$. Further experimental and theoretical development is necessary in order to 
gain a comprehensive understanding of the superconductivity-induced long range magnetic order in the \lco-based 
superconductors. A possible explanation of the trend change of $f_{\rm M}(T)$ at $T_{\rm c}$ could be related 
to charge redistribution between different phases caused by a lowering of the chemical potential upon the opening 
of the superconducting gap in the superconducting phase, a similar mechanism as discussed for the 
superconductivity-induced charge redistributions between different planes in the cuprates \cite{khomskii_PRB46_14245_1992} 
or between different electronic bands in the multi-gap Fe-based superconductors \cite{charnukha_ncomms_2010}. 
In \dlco{N}, as soon as the regions around the SrO-layers turn superconducting, for holes residing in the antiferromagnetic 
regions, it would energetically be favorable to migrate into the ``active'' superconducting layers below $T_{\rm c}$ and thus 
``cleaning up'' the antiferromagnetic layers and leading to a stronger increase of $f_{\rm M}$. This could be possible 
in these systems due to the mesoscopic proximity. Whatever the explanation will prove correctly, the advantage of systems 
as the presented \dlco{N} over the homogeneously doped bulk cuprates is the much higher level of control over the spatial parameters 
in these systems. Further high-resolution transmission electron microscopy and resonant X-ray experiments are necessary 
to verify the correlation of the out-of-plane charge distribution and associated structural distortion \cite{wang_ACS_ApplMaterInterfaces_2016}
with the onset of the superconductivity in  \dlco{N}, in order to shed light on the intriguing interplay between 
superconductivity and long range antiferromagnetic order in the \lco-based superconductors.

  

\begin{acknowledgments}
We gratefully acknowledge S. A. Kivelson, J. Tranquada, C. Bernhard and I. Bozovich for fruitful discussions. 
\end{acknowledgments}

%

\end{document}